\documentclass[12pt]{iopart}

\newcommand{\eqref}[1]{(\ref{#1})}

\usepackage{iopams}
\usepackage{graphicx}
\usepackage{amsfonts}
\usepackage{amsbsy}
\usepackage{amssymb}

\usepackage[top=2cm,bottom=2cm,left=1.5cm,right=1.5cm,columnsep=0.7cm]{geometry}

\usepackage[T1]{fontenc}
\usepackage[utf8]{inputenc}
\usepackage{enumerate}   
\usepackage{bm}

\usepackage{microtype}

\usepackage{color}

\usepackage{array}

\usepackage{citesort}

\begin{document}

\title{Generalized Fluctuation-Dissipation relations holding in non-equilibrium dynamics}

\author{Lorenzo Caprini$^1$}

\address{$^1$ Scuola di Scienze e Tecnologie, Universit\`a di Camerino - via Madonna delle Carceri, 62032, Camerino, Italy}

\ead{ lorenzo.caprini@gssi.it}

\vspace{10pt}
\begin{indented}
\item[]Gennuary 2019
\end{indented}

\begin{abstract}
We derive generalized Fluctuation-Dissipation Relations (FDR) holding for a general stochastic dynamics that includes as subcases both equilibrium models for passive colloids and non-equilibrium models used to describe active particles.
The relations reported here differ from previous formulations of the FDR because of their simplicity: they require only the microscopic knowledge of the dynamics instead of the whole expression of the steady-state probability distribution function that, except for linear interactions, is unknown for systems displaying non-vanishing currents.
From the response function, we can extrapolate generalized versions of the Mesoscopic Virial equation and the equipartition theorem, which still holds far from equilibrium.
Our results are tested in the case of equilibrium colloids described by underdamped or overdamped Langevin equations and for models describing the non-equilibrium behavior of active particles. Both the Active Brownian Particle and the Active Ornstein-Uhlenbeck particle models are compared in the case of a single particle confined in an external potential.
\end{abstract}


\maketitle

\section{Introduction}

The Fluctuation-Dissipation relations (FDR) represent a fundamental topic in statistical physics with a long history. It dates back to the pioneering work of Einstein about the relation between mobility and diffusivity.
Einstein's picture was unified by Kubo~\cite{K57} that through its linear response theory was able to predict the transport coefficients and received an outstanding contribution by the pivotal Onsager's work on reciprocal relations~\cite{O31} holding near the equilibrium.
In these cases, the equilibrium feature of the dynamics and the consequent validity of the detailed balance lead to {\it simple} results that played a crucial role in many areas of physics.

The relation between the response function due to a small perturbation and suitable correlations evaluated in the unperturbed system still represents a fundamental topic to explore non-equilibrium physics leading to the challenging issue of obtaining generalized versions of the FDR holding independently of the detailed balance~\cite{marconi2008fluctuation}.
In the last forty years, several formulations of generalized FDR have been derived using different approaches.
Vulpiani et al.~\cite{FIV90} and Agarwal~\cite{A72} obtained independently generalized FDR for chaotic deterministic and stochastic systems, respectively.
Similar formulations connect the response functions to well-known observables in the framework of stochastic thermodynamics, such as the entropy production~\cite{ss06,seifert2010fluctuation}.
Moreover, both the relations remain, somehow, {\it implicit} since explicitly depend on the steady-state probability distribution function, which is typically unknown for non-equilibrium dynamics.
Successively, path-integral approaches starting from the probability associated with a stochastic trajectory have been employed to derive a new kind of relations, still holding far from the equilibrium.
First examples have been obtained for systems composed by discrete spin variables~\cite{PhysRevE.78.041120} and for Langevin dynamics~\cite{baiesi2009fluctuations,baiesi2010nonequilibrium,baiesi2009nonequilibrium,sarracino2013time,yolcu2017general}, both in the overdamped and underdamped regimes.
Using this approach, Maes et al. focused on the different roles of entropic and frenetic contributions (see, here, for a recent review~\cite{maes}) that distinguish for their parity under time-reversal transformation.
The path-integral technique leads also to another formulation of the FDR connecting the response to a correlation that involves the noise.
The method is known as {\em Malliavin weight sampling}~\cite{malliavin} (see also the Novikov theorem~\cite{n65}) and has been mostly employed in the context of glassy systems to calculate the susceptibility and the effective temperature~\cite{cugliandolo2011effective,CKP94,crisanti2003violation}.
While the Malliavin weight method is particularly efficient and works also for many-body systems, often lacks transparency in the physical meaning of the correlations that are involved in the numerical calculation.

Both the approaches have been recently applied in the context of an emergent class of non-equilibrium dynamics, introduced to describe many biological and physical systems in the framework of Active Matter~\cite{marchetti2013hydrodynamics,bechinger2016active,elgeti2015physics,gompper20202020,shaebani2020computational}.
These systems usually store energy from the environment, for instance through mechanical agents or chemical reactions, to produce directed motion, and represent a good platform to test any version of the generalized FDR~\cite{brady,sarracino2019fluctuation}.
Specifically, the approach of Ref.~\cite{marconi2008fluctuation} has been applied to active matter systems in the limit of small activity (in particular, small persistence time)~\cite{caprini2018linear}. Extending this approach far from equilibrium has the same level of complexity of solving the non-equilibrium active dynamics.
In the same spirit, near-equilibrium FDR have been derived using path-integral techniques~\cite{fodor2016far} leading to a near-equilibrium expression for the susceptibility.
More general results holding also far from equilibrium, both for small and large activities, have been obtained after the generalization of the Malliavin weight sampling procedure to active particle dynamics~\cite{szamel2017evaluating}.
For instance, this technique has been employed to numerically calculate i) the effective temperature of active systems~\cite{berthier2013non,levis2015single,nandi2018effective,cugliandolo2019effective,preisler2016configurational}, with a recent attention to phase-separation~\cite{petrelli2020effective}, and ii) the transport coefficients, such as the mobility, to test an approximated prediction valid at low-density values~\cite{dal2019linear,cengio2020fluctuation} iii) the response function due to a shear flow~\cite{asheichyk2019response}.
Finally, in recent studies based on path-integral approaches, generalized versions of the FDR holding also in far from equilibrium regimes have been reported in the specific case of athermal active particles~\cite{maes2020fluctuating,caprini2020fluctuation}.

In this article, we derive a simple and compact version of the generalized FDR that holds both for equilibrium and non-equilibrium systems.
Our formulation is tested in the case of an equilibrium colloid and an active particle, evaluating both underdamped and overdamped dynamics.
The article is structured as follows: in Sec.~\ref{sec:model}, we introduce a general stochastic model and the notations for the response due to a small perturbative force.
Sec.~\ref{sec:FDR} reports the generalized Fluctuation-Dissipation relations obtained through our approach unveiling its relation with the generalized version of the Mesoscopic Virial equation.
In Sec.~\ref{Sec:Examples}, we test our FDR for the underdamped and overdamped dynamics of equilibrium colloids and for two popular non-equilibrium models introduced to describe the behavior of active particles.
Finally, in Sec.~\ref{sec:comparison}, we compare our FDR to earlier relations, while in the last section we present the conclusions.
\section{The Response function due to a small perturbation}\label{sec:model}

To define the response function and successively derive our version of the FDR, we introduce a general stochastic dynamics which describes the evolution of a set of $N$ variables, namely $\mathbf{x}=(x_1, x_2, ... , x_N)$.
The set of stochastic differential equations, of which we refer to as the {\it unperturbed dynamics}, is the following:
\begin{equation}
\label{eq:dynamics}
\dot{x}_i = \mathcal{F}_i (\mathbf{x}, t) + \sigma_{ij}\cdot \xi_j \,,
\end{equation}
where we have adopted the Einstein summation convention.
The term $\mathcal{F}_i (\mathbf{x}, t)$ contains all the deterministic contributions ruling the dynamics of $x_i$ and in the following will be denoted as a {\it{force}}.
This is a general function that could depend on the whole set of the state variables and could even contain an explicit dependence on the time, $t$.
This choice of the force allows us to describe both equilibrium systems characterized by Boltzmann distributions and non-equilibrium systems with non-vanishing steady-state currents induced by $\mathcal{F}_i (\mathbf{x}, t)$.
The general dynamics~\eqref{eq:dynamics} includes a broad range of non-equilibrium models that have been largely employed to describe systems of biological and/or technological interest, for instance in the context of active matter. These examples will be explicitly discussed in the final part of Sec.~\ref{Sec:Examples}.
The term $\xi_j$ is a white noise with zero average and unit variance, such that
\begin{equation*}
\langle \xi_j(t) \cdot \xi_i(s)\rangle=2\delta_{ji} \delta(t-s) \,,
\end{equation*}
where $\delta_{ij}$ is the Kronecker function and $\delta(t-s)$ is the Dirac-$\delta$ function.
Finally, the term $\sigma_{ij}$ is a general matrix that determines the amplitudes of each noise term (and that could also be  non-symmetric).
Its square gives rise to the diffusion matrix, $D_{ij}=\sigma_{ik}\sigma_{kj}$. 
The dynamics~\eqref{eq:dynamics} has a very general form and could include also deterministic variables if the matrix $\boldsymbol{\sigma}$ is singular, as in the usual case of a particle described in terms of position and velocity.
From now, we choose $\boldsymbol{\sigma}$ as a general matrix with constant elements, restricting our analysis to the case of additive noise and, thus, excluding any multiplicative dynamics.

Perturbing the dynamics~\eqref{eq:dynamics} means adding a force $h_i(t)$ smaller than the other force contributions so that the perturbed variables, $x_i^h$, which will be denoted by the superscript $h$, evolves as:
\begin{equation}
\label{eq:dynamics_perturbed}
\dot{x}_i^h =  \mathcal{F}_i (\mathbf{x}^h, t) + \sigma_{ij}\, \xi_j +  h_i(t)\,.
\end{equation}
As a result, the set of perturbed variables $\mathbf{x}^h$ deviate from the unperturbed set of variables $\mathbf{x}$ by $\delta x_j(0)$.
We choose $h_j(t)=\delta x_j(0) \delta(t)$, where $\delta(t)$ is the Dirac function and $\delta x_j(0)=x_j^h - x_j$ is the deviation of the perturbed variable from the unperturbed one.
The response function, due to this small perturbation, of a general observable $A(\mathbf{x}, t)$, that depends on the whole set of variables and explicitly on the time, is defined as:
\begin{equation}
\label{eq:response_def}
\mathcal{R}_{A, x_j} (t) =  \frac{\langle A(\mathbf{x}(t))\rangle^h - \langle A(\mathbf{x}(t)) \rangle}{\delta x_j(0)} \equiv \frac{ \delta\langle A(\mathbf{x}^h)\rangle}{\delta h_j(0)}\biggr|_{h_n=0}  \,,
\end{equation}
where $\delta/\delta h_j$ is the functional derivative with respect to $h_j$ calculated at $h_j=0$.
The response function can be calculated numerically through Eq.~\eqref{eq:response_def} that requires the knowledge of the perturbed dynamics.

\section{Generalized Fluctuation-Dissipation Relations}\label{sec:FDR}

As mentioned in the introduction, the idea of expressing the response due to a small perturbation in terms of unperturbed correlations has a long history.
Here, we report a version of the generalized FDR holding independently of the presence of non-vanishing currents and, thus, valid also for non-equilibrium dynamics.
This version of the FDR does not depend explicitly on the probability distribution function but requires the explicit knowledge of the microscopic dynamical details, i.e. the knowledge of $\sigma_{ij}$ and $\mathcal{F}_{i}(\mathbf{x})$.
Using a path-integral formalism and assuming the stationarity of the time-properties, the response function, $\mathcal{R}_{A, x_j} (t)$, associated to the dynamics~\eqref{eq:dynamics} and~\eqref{eq:dynamics_perturbed}, can be expressed as: 
\begin{equation}
\label{eq:FDR_A_general}
\mathcal{R}_{A, x_j} (t) = -\frac{1}{2} D_{jm}^{-1} \left[ \left\langle A(t) \mathcal{F}_m(0) \right\rangle + \frac{d}{dt}\left\langle A(t)  x_m(0) \right\rangle  \right] \,,
\end{equation}
where the dependence on $\mathbf{x}$ in $A(t)=A(\mathbf{x}(t))$ and $\mathcal{F}(\mathbf{x}(t))=\mathcal{F}(t)$ has been omitted for notational convenience and the average on the right-hand side, $\langle \cdot\rangle$, is calculated through the unperturbed dynamics.
Further details about the derivation of Eq.~\eqref{eq:FDR_A_general} are reported in~\ref{app:1}.
In addition, we remark that these exact relations are not simply expressed by the temporal correlation between $A$ and another observable at variance with equilibrium. Our FDR explicitly contains the time-derivative of the observable $A$ and the state variable.
Fixing $A(\mathbf{x})=x_k$ allows us to consider the expression for the response matrix of element $\mathcal{R}_{x_k, x_j}(t)$.
In this specific case, the relation~\eqref{eq:FDR_A_general} further simplifies leading to the following steady-state expression that is derived in ~\ref{app:B}:
\begin{equation}
\label{eq:FDR_x_general}
\mathcal{R}_{x_k, x_j} (t) = -\frac{1}{2} D_{jm}^{-1} \left[ \langle x_k(t) \mathcal{F}_m(0) \rangle +  \langle \mathcal{F}_k(t) x_m(0) \rangle \right] \,.
\end{equation}
The relations~\eqref{eq:FDR_A_general} and~\eqref{eq:FDR_x_general} are general for every equilibrium and non-equilibrium dynamics of the form~\eqref{eq:dynamics}.
We remark that the response matrix elements assumes a form particularly simple: they are expressed as temporal correlations between the state variables and the forces that rule the dynamics, combined by the elements of the diffusion matrix.
In the diagonal case, $D_{ij}=\delta_{ij} D_j$, in particular, Eq.~\eqref{eq:FDR_x_general} turns to be:
\begin{equation}
\label{eq:FDR_x_general_diagonal}
\mathcal{R}_{x_k, x_j} (t) = -\frac{1}{2 D_j}  \left[ \langle x_k(t) \mathcal{F}_j(0) \rangle +  \langle \mathcal{F}_k(t) x_j(0) \rangle \right] \,.
\end{equation}
so that each element of the response matrix is given by the sum of two correlations: i) the time correlation between the observed variable and the force ruling the dynamics of the perturbed variable and ii) the same correlation with swapped times.
In Appendices~\ref{app:1} and~\ref{app:B}, the relations~\eqref{eq:FDR_A_general} and~\eqref{eq:FDR_x_general} are evaluated assuming to deal with a dynamics where the detailed balance holds, showing their consistence with known equilibrium results.

In Sec.~\ref{Sec:Examples}, we consider equilibrium and non-equilibrium dynamics to test our exact relations.
In particular, we evaluate the case of passive equilibrium colloids both in the overdamped and underdamped regimes showing the agreement with other versions of equilibrium FDR.
Successively, we apply our generalized FDR to systems of active particles to check our relations in non-equilibrium models.

\subsection{Response Function, Generalized Equipartion theorem and Mesoscopic Virial equation}

In this section, we show that we can extract a relation between suitable steady-state correlations from
the response matrix and our version of generalized FDR.
Indeed, the response matrix at the perturbation time, $R_{x_k x_j}(0)$, is not arbitrary because of its definition~\eqref{eq:response_def}. In particular, $R_{x_k x_j}(0)=\delta_{jk}$ since its diagonal elements are unitary, while the cross elements vanish because of the causality condition.
Therefore, evaluating the FDR, Eq.~\eqref{eq:FDR_x_general}, at the perturbation time, leads to the following tensorial relation:
\begin{equation}
\label{eq:generalized_equipartition}
2 \delta_{jk} = - D^{-1}_{j m} \left[\langle x_k \mathcal{F}_m\rangle + \langle \mathcal{F}_k x_m \rangle \right] \,.
\end{equation}
Equation~\eqref{eq:generalized_equipartition} establishes a set of exact relations between special equal-time averages that are functions of the state variables of the dynamics through ${x}_k$ and $\mathcal{F}_m(\mathbf{x})$ (that is expressed as 
$\mathcal{F}_m$ for notational convenience).
These relations hold for both equilibrium and non-equilibrium systems and their physical interpretation will be clarified in the explicit examples reported in Sec.~\ref{Sec:Examples}. 
In particular, we anticipate that Eqs.~\eqref{eq:generalized_equipartition} represent a generalization of the Equipartition theorem and the Mesoscopic Virial equation.
Despite these equations can be obtained via other methods in many interesting cases, we stress that they are also contained in our version of the FDR, from which their derivation is straightforward. 

In the case of diagonal diffusion, such that $D_{ij}=D_j \delta_{ij}$, Eq.~\eqref{eq:generalized_equipartition} assumes a simpler form.
These relations provide general constraints for the matrix, $\mathcal{M}$, of elements $\mathcal{M}_{jk}=\langle x_j \mathcal{F}_k\rangle$, involving the steady-state correlation between the state variable $x_j$ and the deterministic force that determines the evolution of $x_k$.
The diagonal elements of $\mathcal{M}$ satisfy:
\begin{equation}
 D_j = -\langle x_k \mathcal{F}_j\rangle \delta_{jk}\,,
\end{equation}
that can be interpreted as a generalized version of the equipartition theorem as illustrated in Sec.~\ref{Sec:Examples}.
Instead, the off-diagonal elements of $\mathcal{M}$ are constrained by the following relation:
\begin{equation}
\label{eq:outdiagonal}
  \langle x_k \mathcal{F}_j\rangle  = -  \langle x_j \mathcal{F}_k\rangle \,,
\end{equation}
with $k \neq j$.
Therefore, the matrix $\mathcal{M}$ is anti-symmetric.
As we can see in Sec.~\ref{Sec:Examples}, the relation~\eqref{eq:outdiagonal} represents a generalized version of the Mesoscopic Virial equation.
These relations have been derived for the specific case of a particle, following the Langevin dynamics, by Falasco et al.~\cite{falasco2016mesoscopic} using a different approach while, here, are extended to a more general dynamics and connected to our version of the generalized FDR.

\section{Examples}\label{Sec:Examples}

\subsection{Passive Colloidal dynamics}

To test our general results, we start by considering the equilibrium dynamics describing the motion of a passive colloidal particle in a solvent. In this case, the generalized FDR need to be consistent with the well-known FDR holding at equilibrium. 
%
Specifically, assuming that the colloid is in equilibrium with the solvent at temperature, $T$, and neglecting hydrodynamics interactions, the dynamics for the particle position, $\mathbf{x}$, and the particle velocity, $\mathbf{v}$, reads:
\begin{eqnarray}
\label{eq:dynamics_passivecolloids}
\dot{\mathbf{x}}&&=\mathbf{v} \\
m\dot{\mathbf{v}}&&=-\gamma\mathbf{v} + \mathbf{F} + \sqrt{2\gamma T} \boldsymbol{\xi}\,,
\end{eqnarray}
where $m$ is the mass of the colloid, $\gamma$ the drag coefficient and $T$ the solvent temperature that satisfies the Einstein relation with the diffusion coefficient, $\gamma D_t=T$.
The term $\mathbf{F}$ accounts for external forces due to a potential, such that $\mathbf{F}=-\nabla U(\mathbf{x})$, while the term $-\gamma\mathbf{v}\equiv \mathbf{F}_s$ is the Stokes force proportional to the velocity. This term balances the injection of energy due to the collisions of the solvent particles that are modeled through a white noise. The diffusion matrix is diagonal, such that $\sigma_{ij}=\delta_{ij}\sqrt{2T\gamma}/m$, since there are no temperature gradient.
In this case, the response of an observable $A(\mathbf{v}, \mathbf{x})$ due to the additional perturbative force $h_j$ reads:
\begin{equation}
\label{eq:numericalresponse_u}
\mathcal{R}_{A, v_j}(t)= \frac{\langle A(\mathbf{x}(t))\rangle^h - \langle A(\mathbf{x}(t)) \rangle}{\delta v_j^h(0)} = m\frac{ \delta\langle A(\mathbf{x}^h)\rangle}{\delta h_j}\biggr|_{h_j=0}\,,
\end{equation} 
where we remind that the average $\langle \cdot \rangle^h$ is realized through the perturbed measure and the Latin indices are used to denote the Cartesian components of the vectors, here and in the next examples.
Thus, in this case, the set of variables is composed of the $d$ Cartesian components of position and velocity where $d$ is the dimension of the system.

Applying the general formula~\eqref{eq:FDR_x_general} with the dynamics~\eqref{eq:dynamics_passivecolloids}, i.e. replacing $\mathcal{F}_j= - \gamma v_j /m - \nabla_{x_j} U/m$, leads to the following result for the response matrix:
\begin{equation}
\label{eq:R_v_underdampedpassive}
\mathcal{R}_{v_k, v_j} (t) = \frac{m}{ T }  \langle v_k(t) v_j(0)  \rangle +  \frac{m}{2 T \gamma } \left[ \langle v_k(t) \nabla_{x_j}U(0) \rangle +  \langle \nabla_{x_k}U(t) v_j(0) \rangle \right] \,.
\end{equation}
Since, by definition, the system is in equilibrium, the detailed balance holds and we can further manipulate Eq.~\eqref{eq:R_v_underdampedpassive} by using the time-reversibility of the steady-state correlation such that $\langle v_k(t) \nabla_{x_j}U(0) \rangle = - \langle \nabla_{x_j} U(t) v_k(0) \rangle$. 
In addition, we can use the symmetry among different Cartesian components, such that $\langle \nabla_{x_k}U(t) v_j(0) \rangle = \langle \nabla_{x_j}U(t) v_k(0) \rangle$, that is valid for central potentials.
Using these properties, the square brackets in Eq.~\eqref{eq:R_v_underdampedpassive} vanish and we obtain the well-known equilibrium result, $\mathcal{R}_{v_k, v_j} (t) = \frac{m}{ T }  \langle v_k(t) v_j(0) \rangle$.
Finally, choosing $A(\mathbf{x}, \mathbf{v})=x_k$ one can calculate the cross terms of the response matrix (coupling position and velocity) starting from Eq.~\eqref{eq:FDR_A_general}:
\begin{equation}
\label{eq:resp_xv_underdampedpassive}
\mathcal{R}_{x_k, v_j} (t) = \frac{m}{ 2T }  \langle x_k(t) v_j(0)  \rangle +  \frac{m}{2 T \gamma } \langle x_k(t) \nabla_{x_j}U(0) \rangle  - \frac{m^2}{2T\gamma}\langle v_k(t) v_j(0) \rangle \,.
\end{equation}
We observe that in equilibrium systems the relations~\eqref{eq:resp_xv_underdampedpassive} vanish term by term except for $j=k$ where only the second and the third terms survive.
Using the time-reversibility, the equation of motion and tricks similar to those employed to manipulate Eq.~\eqref{eq:R_v_underdampedpassive}, also in this case, we can recover the well-known result holding in equilibrium, that is $\mathcal{R}_{x_k, v_j} (t) = \frac{m}{ T }  \langle x_k(t) v_j(0)  \rangle$. 

\subsubsection{Overdamped Dynamics}

The dynamics of an equilibrium colloidal particle is often described by an overdamped stochastic differential equation for the position, $\mathbf{x}$, because the inertial forces play a negligible role.
In this case, the evolution of each colloid is described by the following equation:
\begin{equation}
\label{eq:dynamics_passivecolloidoverdamped}
\gamma\dot{\mathbf{x}} =- \nabla U + \sqrt{2\gamma T} \boldsymbol{\xi} \,.
\end{equation}
In the overdamped case, one can calculate the response function of an observable $A(\mathbf{x})$ perturbing directly the particle position that is a noisy variable, i.e. the dynamics~\eqref{eq:dynamics_passivecolloidoverdamped}.
Therefore, the response function, $\mathcal{R}_{A, x_j}(t)$, is defined as:
\begin{equation}
\label{eq:numericalresponse_o}
\mathcal{R}_{A, x_j}(t)=\frac{\langle A(\mathbf{x}(t))\rangle^h - \langle A(\mathbf{x}(t)) \rangle}{\delta x_j(0)}= \gamma\frac{ \delta\langle A(\mathbf{x}^h)\rangle}{\delta h_j}\biggr|_{h_n=0} \,.
\end{equation} 
Now, the set of variables involved in the FDR contains only the $d$ Cartesian components of the position. 
After identifying $\mathcal{F}_k=-\nabla_{x_k} U/\gamma$ and $\sigma_{jk}=\delta_{jk}\sqrt{2T/\gamma}$, we can apply formula~\eqref{eq:FDR_x_general} so that the response function reads:
\begin{equation}
\label{eq:R_x_overdampedpassive}
\mathcal{R}_{x_k, x_j} (t) = \frac{1}{2 T  } \left[ \langle x_k(t) \nabla_{x_j}U(0) \rangle +  \langle \nabla_{x_k}U(t) x_j(0) \rangle \right] \,.
\end{equation}
The well-known FDR can be recovered again by using the time-reversibility so that $ \langle \nabla_{x_k}U(t) x_j(0)\rangle =  \langle x_j(t) \nabla_{x_k}U(0)  \rangle$. The absence of currents also implies that the system is invariant for changes of Cartesian components so that $\langle x_j(t) \nabla_{x_k}U(0)  \rangle=\langle x_k(t) \nabla_{x_j}U(0)  \rangle$.
In this way, Eq.~\eqref{eq:R_x_overdampedpassive} reduces to the well-known equilibrium result,
$\mathcal{R}_{x_k, x_j} (t) = \frac{1}{T } \langle x_k(t) \nabla_{x_j} U(0) \rangle$.

\subsubsection{Generalized Virial equation and Equipartition theorem}

In the case of a passive underdamped colloid, following the dynamics~\eqref{eq:dynamics_passivecolloids}, the relation~\eqref{eq:generalized_equipartition} turns to be:
\begin{equation}
\label{eq:passive_colloidal_equip}
 m\langle v_k v_j \rangle + \frac{m}{\gamma}\langle v_j \nabla_{x_k} U \rangle  =  T \delta_{kj} \,.
\end{equation}
The diagonal elements of this relation for $j=k$ can be further manipulated since $\langle v_k \nabla_{x_k} U \rangle = 1/t_f \int_{0}^{t_f} d/dt U(\mathbf{x}(t)) dt = [U(\mathbf{x}(t_f))-U(\mathbf{x}(0))]/t_f$, that is the potential energy difference from the initial and the final state. Since this term gives a negligible contribution in the steady-state ($t_f \to \infty$), the relation~\eqref{eq:passive_colloidal_equip} trivially holds and states that $\langle v^2_k\rangle= T$, in agreement with the equilibrium distribution $\propto \exp{\left(- U/T - m \sum_k v_k^2/2 T\right) }$.
The equation for the off-diagonal terms implies that $\langle v_k v_j \rangle =- \frac{1}{\gamma}\langle v_j \nabla_{x_k} U \rangle$.
where each correlation is zero.
The cross-correlation of the response matrix, coupling position and velocity, i.e. Eq.~\eqref{eq:resp_xv_underdampedpassive} at the perturbation time, leads to the following relation:
\begin{equation}
\label{eq:passive_colloidal_virial}
 \langle x_j v_k \rangle + \langle x_j \nabla_{x_k} U \rangle  -  m\langle v_j v_k \rangle = 0 \,.
\end{equation}
if $j=k$, the first term vanishes because is a boundary term, such that $\langle x_k v_k \rangle = 1/t_f \int_{0}^{t_f} d/dt \mathbf{x}(t)^2/2 dt = [x^2(t_f)-x^2(0)]/(2t_f)$, that is irrelevant for large times ($t_f \to \infty$).
Moreover, the second term of Eq.~\eqref{eq:passive_colloidal_virial} can be identified as the virial pressure and is related to the kinetic energy by this formula.
If $U$ contains also an interacting potential with other colloidal particles, this equation is nothing but the Virial mesoscopic equation, that has been derived in Ref.~\cite{falasco2016mesoscopic} using a different method. 

In the case of passive overdamped colloids following the dynamics~\eqref{eq:dynamics_passivecolloidoverdamped}, we can apply the relation~\eqref{eq:generalized_equipartition}, obtaining:
\begin{equation}
\label{eq:overdamped_virialequation}
\delta_{jk} T=  \langle x_j \nabla_{x_k} U \rangle \,.
\end{equation}
Equation~\eqref{eq:overdamped_virialequation} can be derived directly from Eqs.~\eqref{eq:passive_colloidal_equip} and~\eqref{eq:passive_colloidal_virial} assuming that the inertial time $m/\gamma$ is small, just by considering the different contributions in powers of $m/\gamma$. Again, this equation holds since the equilibrium distribution is $\propto \exp{\left(- U/T  \right)}$ and states that the Virial pressure is determined by the solvent temperature.

\subsection{Self-Propelled Particles}

Active particles are usually described by stochastic equations that resemble those of passive colloids moving in viscous solvents except for the addition of a time-dependent stochastic force called self-propulsion or simply active force.
Usually, the active force is chosen to reproduce the typical time-persistence of the active trajectory at a coarse-grained level that neglects its mechanical or chemical origin (which depends on the system under consideration). 
This force, except for a few special cases, breaks the detailed balance~\cite{marconi2017heat,dabelow2020irreversibility} condition producing a non-vanishing entropy production~\cite{caprini2018comment,caprini2019entropy,mandal2017entropy,shankar2018hidden,chaki2019effects,dabelow2020irreversible}.
Therefore, active dynamics are good platforms to evaluate generalized FDR in far-equilibrium systems.

The most popular and simple models to reproduce the self-propulsions through a stochastic process are the Active Brownian Particles (ABP) dynamics~\cite{buttinoni2013dynamical,solon2015pressure,fily2019self,stenhammar2014phase,farage2015effective,das2018confined,digregorio2018full,mandal2019motility,caprini2020hidden} and the Active Orstein Uhlembeck particles (AOUP) one~\cite{wittmann2017effective,caprini2018active,maggi2017memory,woillez2020active,dabelow2019irreversibility,berthier2019glassy,martin2020statistical}. Both have been used to reproduce the non-equilibrium phenomenology of self-propelled particles. 
In the ABP case, the self-propulsion force, $\mathbf{f}_a$ has a constant modulus and reads:
\begin{equation*}
\mathbf{f}^a=\gamma v_0 \mathbf{n} \,,
\end{equation*}
being $v_0$ the swim velocity induced by the self-propulsion and $\gamma$ the viscous solvent. The term $\mathbf{n}_i=(\cos\theta_i, \sin\theta_i)$ is a unit vector representing the particle orientation since $\theta$ is the orientational angle that evolves via a Brownian motion:
\begin{equation*}
\dot{\theta} = \sqrt{2 D_r} \xi \,,
\end{equation*}
where $D_r$ is the rotational diffusion coefficient and $\xi$ is a white noise with zero average and unit variance.
According to the AOUP scheme, the self-propulsion of each particle is described by a vectorial Ornstein-Uhlenbeck process:
\begin{equation}
\label{eq:AOUP}
\tau\dot{\mathbf{f}}^a = - {\mathbf{f}}^a  + \gamma v_0\sqrt{\tau} \, \boldsymbol{\eta} \,,
\end{equation}
where $\tau$ is the persistence time of the process, $\boldsymbol{\eta}$ is a vector of white noises with zero average and unit variance, and the other parameters have been already introduced.
Here, the term $v_0^2 \gamma^2$ is the variance of the self-propulsion whose square root also represents the average value of its modulus, which, thus, provides the same average swim velocity of the ABP.
Despite the different shapes of ABP and AOUP models, they share important time-dependent properties that are considered responsible for their common phenomenology.
Even if many experimental systems of active matter have microscopic sizes~\cite{bechinger2016active} and usually move in environments with large viscosity (in such a way that inertial forces are negligible), recently, the effects of inertia~\cite{lowen2020inertial} have been highlighted in many experimental active systems, such as vibro-robots~\cite{scholz2018inertial}, Hexbug crawlers and camphor surfers~\cite{leoni2020surfing} and 
vibration-driven granular particles~\cite{puglisi2012structure,weber2013long,dauchot2019dynamics} (in the granular case, the response function has been also calculated experimentally~\cite{gnoli2014nonequilibrium}). 
To include the active force in these physical systems, the active Langevin model has been introduced~\cite{scholz2018inertial,um2019langevin,lowen2020inertial,vuijk2020lorentz,sprenger2021time,caprini2021inertial} so that the equation of motion of the active particle is described by its position, $\mathbf{x}$, and velocity, $\mathbf{v}$: 
\begin{eqnarray}
\label{eq:dynamics_activecolloids}
\dot{\mathbf{x}} &&= \mathbf{v}\\
m\dot{\mathbf{v}} &&= -\gamma \mathbf{v} + \mathbf{F} + \mathbf{f}^a + \sqrt{2T\gamma} \,\mathbf{w} \,, 
\end{eqnarray}
where $\mathbf{w}$ is a white noise vector with zero average and unit variance, $\mathbf{f}^a$ is the active force discussed above, and the other terms have been already introduced below Eq.~\eqref{eq:dynamics_passivecolloids}.

Both for ABP and AOUP active forces, we can obtain a generalized FDR for the elements of the response matrix, defined by Eq.~\eqref{eq:numericalresponse_u}, by applying the formula~\eqref{eq:FDR_x_general} with $\mathcal{F}_j = - \gamma v_j /m - \nabla_{x_j} U/m + \mathrm{f}^a_j/m$ and $\sigma_{ij}=\delta_{ij}\sqrt{2T\gamma}/m$: 
\begin{eqnarray}
\label{eq:active_underdamped_x}
\mathcal{R}_{v_k, v_j}(t) =&& \frac{m}{T } \left\langle v_k(t) v_j(0) \right\rangle + \frac{m}{2 T \gamma} \left(\left\langle v_k(t) \nabla_{x_j}U(0) \right\rangle + \left\langle \nabla_{x_k} U(t) v_j(0) \right\rangle  \right) \nonumber\\
&&-\frac{m}{2T \gamma} \left(\left\langle v_k(t) \mathrm{f}_j^a(0) \right\rangle + \left\langle \mathrm{f}^a_k(t) v_j(0) \right\rangle  \right) \,.
\end{eqnarray}
The first term in the right-hand-side of Eq.~\eqref{eq:active_underdamped_x} represents the response in the equilibrium regimes, i.e. for $v_0 \to 0$ that corresponds to the well-known results for passive Brownian particles reported in Eq.~\eqref{eq:R_v_underdampedpassive}.
The second and third terms are non-equilibrium contributions of the response that exactly balance at equilibrium where the detailed balance holds since these terms are odd under time-reversal symmetry.
The fourth and fifth terms of the second line, instead, are truly non-equilibrium contributions that explicitly contain the self-propulsion force.
We remark that we do not need to specify the parity under time-reversal transformation of the active force since this information is not required for the calculation of the response.
In a similar way, we can calculate the cross elements of the response matrix, choosing $A(\mathbf{x}, \mathbf{v})=x_k$ in Eq.~\eqref{eq:active_underdamped_x}, obtaining:
\begin{equation}
\label{eq:active_underdamped_x_cross}
\mathcal{R}_{x_k, v_j} (t) = \frac{m}{ 2T }  \langle x_k(t) v_j(0)  \rangle +  \frac{m}{2 T \gamma } \langle x_k(t) \nabla_{x_j}U(0) \rangle -  \frac{m}{2 T \gamma } \langle x_k(t) \mathrm{f}^a_j(0) \rangle  - \frac{m^2}{2T\gamma}\langle v_k(t) v_j(0) \rangle \,.
\end{equation}
We remark that both Eqs.~\eqref{eq:active_underdamped_x} and~\eqref{eq:active_underdamped_x_cross} hold far from the equilibrium without restriction in the parameters of the self-propulsion, at variance with other approaches where the active force is considered as a small perturbation~\cite{sharma2016communication,merlitz2018linear}.

We check our theoretical results by studying the elements of the response matrix, $\mathcal{R}_{v_j,v_j}(t)$ and $\mathcal{R}_{v_j, x_j}(t)$, confining the system through a linear and a quartic potential in two dimensions. 
The time is calculated in unit of $t^*=m/\gamma$.
In both cases, the cross elements of the response function ($\mathcal{R}_{v_k,v_j}(t)$ and $\mathcal{R}_{x_k,v_j}(t)$ with $k \neq j$) are zero for symmetric arguments:
indeed, each correlation appearing in the FDR should be invariant under the transformation $x_j \to -x_j$, $v_j \to -v_j$ and $f^a_j \to -f^a_j$ at fixed $j$.
Since all the terms appearing in the cross elements of Eq.~\eqref{eq:Resp_x_overdamped_active} are odd under this transformation, the only possibility is that $R_{v_k , v_j}(t)=R_{x_k , v_j}(t)=0$  if $k \neq j$.

\begin{figure}[!t]
\centering
\includegraphics[width=0.99\linewidth,keepaspectratio]{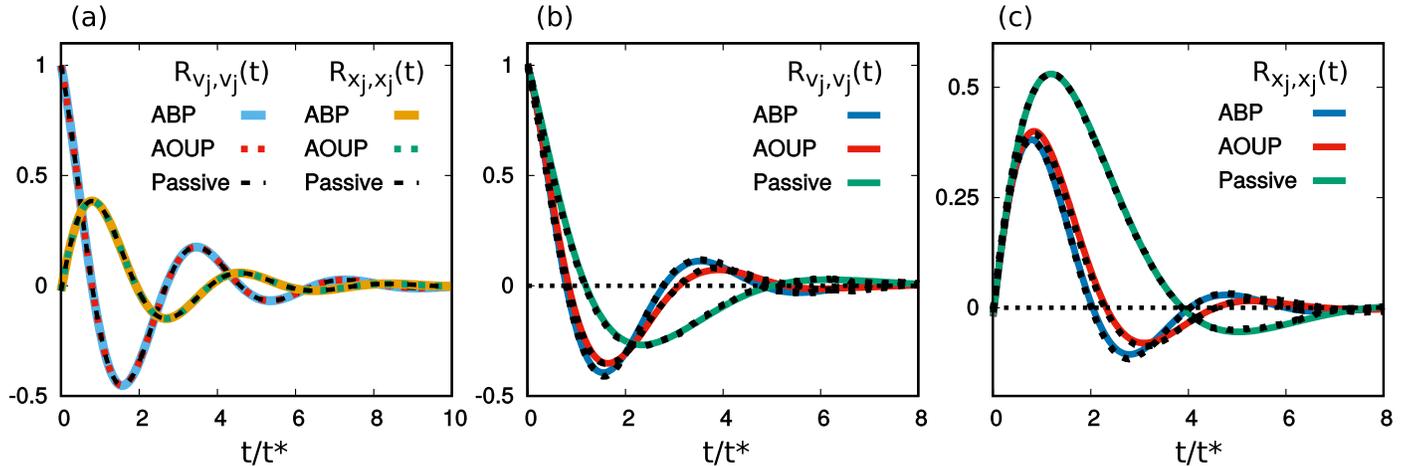}
\caption{\label{fig:under}
Elements of the response matrix, $R_{v_j, v_j}(t/t^*)$ and $R_{x_j, v_j}(t/t^*)$, obtained numerically using the definition~\eqref{eq:numericalresponse_u} (colored lines) and using the relation~\eqref{eq:active_underdamped_x} and~\eqref{eq:active_underdamped_x_cross} (dashed black lines). 
In panel~(a), $R_{v_j v_j}(t/t^*)$ and $R_{x_j v_j}(t/t^*)$ are studied for the harmonic confinement, $U(\mathbf{x})=k\,|\mathbf{x}|^2/2$, for passive, AOUP and ABP particles. The shapes of $R_{v_j, v_j}(t/t^*)$ and $R_{x_j, v_j}(t/t^*)$ are reported for a single set of parameters (also in the active cases) since the only dependence on them is contained in $t^*=m /\gamma$ and $\omega^2=k/m- (t^*)^{-2}/4$. In addition, we have simply plotted the passive profile, $R(t)=\exp{(-t/t^*)}\left[\cos{(\omega t)} - \sin{(\omega t)}/(2t^*\omega)\right]$, for the passive system.
Panels~(b) and~(c) report $R_{v_j, v_j}(t/t^*)$ and $R_{x_j, v_j}(t/t^*)$, respectively, for a system confined through a quartic potential, $U(\mathbf{x})=k\,|\mathbf{x}|^4/4$.  
Here, the additional dotted black lines are eye-guides 
The other parameters are $k=3$, $\gamma=1$, $T=1$, $\tau=1$ and $v_0=1$. 
}
\end{figure}

The response in the harmonic passive case, with $U(\mathbf{x})=k\, |\mathbf{x}|^2/2$ (where $k$ is the potential constant), can be analytically solved because the velocity correlation appearing in the FDR can be calculated as a function of $t$ and depends on the inertial time $t^*=m/\gamma$ and on the frequency $\omega^2=k/m - (t^*)^{-2}/4$, as known in the literature.
As shown in Fig.~\ref{fig:under}~(a), the profile of $\mathcal{R}_{v_j, v_j}(t)$ and $\mathcal{R}_{x_jv_j}(t)$ both for the AOUP and the ABP dynamics remains the same as a result of the linearity of the force. This occurs even if, in both cases, the functional form of the FDR changes because of the non-vanishing time correlation between $x_k$ and $\mathrm{f}_k^a$.
Fig.~\ref{fig:under}~(b) reports a similar study when passive or active dynamics are confined by the quartic potential, $U(\mathbf{x})=k\, |\mathbf{x}|^4/4$.
In this case, there are no analytical solutions for $\langle v_j(t) v_j(0)\rangle$ (and for the other correlations) neither in the passive nor in the active cases, because of the non-linearity of the dynamics.
Therefore, the validity of the FDR is checked numerically by comparing the elements of $\mathcal{R}$ calculated by their definition~\eqref{eq:numericalresponse_u} and by the FDR, and shows a good agreement.
Besides, the functional forms of $\mathcal{R}_{v_j, v_j}(t)$ and $\mathcal{R}_{x_j, v_j}(t)$ in the active cases (both ABP and AOUP) display more pronounced oscillations that also occur for smaller times with respect to the passive profile of the response.
Additionally, the difference between AOUP and ABP dynamics appears only in the limit $v_0^2 \gg T$ (and increases with the growth of $v_0^2/T$, while in the opposite limit (not shown) the AOUP and ABP responses become equal to each other before converging to the passive profiles when the active force is negligible.

\subsubsection{Overdamped Dynamics for Self-propelled Particles}

We also study the active dynamics directly in the overdamped regime.
Since the inertial forces are usually negligible in many experimental active systems, the overdamped limit has been largely employed in most of the numerical studies about active matter and, thus, deserves particular attention.
The resulting dynamics is a stochastic differential equation for the particle position $\mathbf{x}$: 
\begin{equation}\label{eq:motion}
\gamma\dot{\mathbf{x}} =\mathbf{F} + \mathbf{f}^a +\sqrt{2 T  \gamma} \mathbf{w} \,. 
\end{equation}
Once the velocities have been eliminated, the positions evolve through a stochastic dynamics and we can calculate the response function, defined by Eq.~\eqref{eq:numericalresponse_o}.
Taking $\sigma_{ij}=\delta_{ij} \sqrt{D_t}$ and $\mathcal{F}_j = - \nabla_{x_j} U/\gamma + \mathrm{f}^a_j/\gamma$, we apply the general Eq.\eqref{eq:FDR_x_general} to calculate $R_{x_k x_j}$ obtaining the following FDR for the elements of the response matrix:
\begin{equation}
\label{eq:Resp_x_overdamped_active}
\mathcal{R}_{x_k, x_j}(t) =   \frac{1}{2 T } \left(\left\langle x_k(t) \nabla_{x_j} U(0) \right\rangle + \left\langle \nabla_{x_k}U(t) x_j(0) \right\rangle  \right) - \frac{1}{2T } \left(\left\langle x_k(t) \mathrm{f}^a_j(0) \right\rangle + \left\langle \mathrm{f}^a_k(t) x_j(0) \right\rangle  \right) \,.
\end{equation}
The first and the second terms are the equilibrium-like contributions of the response that coincides only if the detailed balance holds and that are otherwise different.
Instead, the second and third terms are the non-equilibrium contributions involving the time-correlation of active force and position that disappears in the equilibrium limit, $v_0\to0$.

\begin{figure}[!t]
\centering
\includegraphics[width=0.99\linewidth,keepaspectratio]{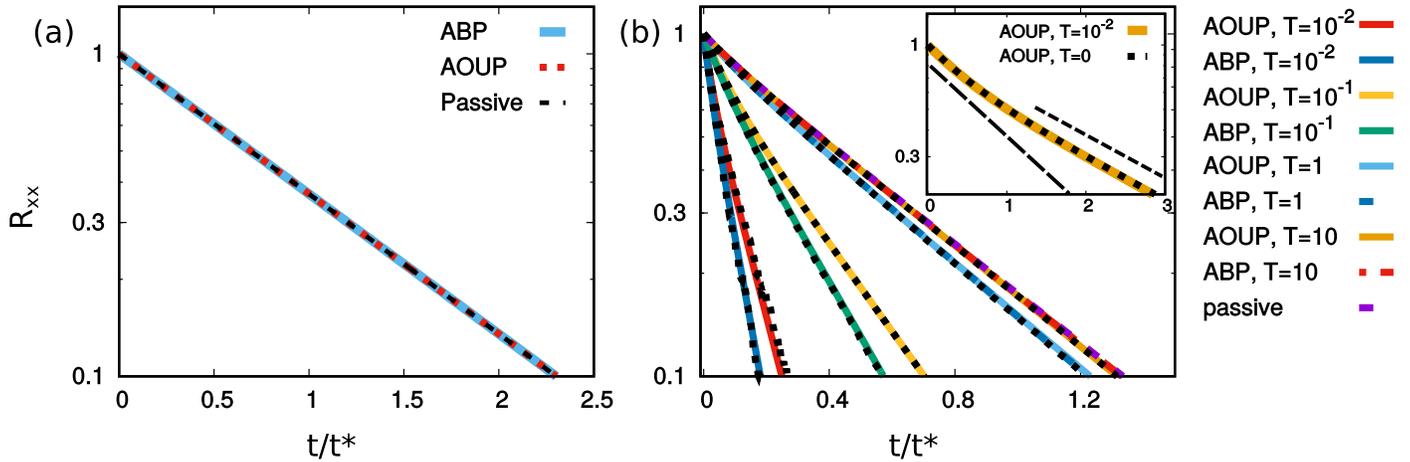}
\caption{\label{fig:over}
Response function, $R_{xx}(t/t^*)$, obtained numerically using the definition~\eqref{eq:numericalresponse_o} (colored lines) and using the relation~\eqref{eq:Resp_x_overdamped_active} (dashed black lines). 
In panel (a), $R_{xx}(t/t^*)$ is studied for the harmonic confinement, $U(\mathbf{x})=k\,|\mathbf{x}|^2/2$, in the cases of passive, AOUP and ABP particles. In this case, we report the expression for a single set of parameters (also in the active cases) since the only dependence on them is contained in $t^*=t/\gamma$. In addition, we have simply plotted the exponential profile, $\exp{(-t/t^*)}$, for the passive system.
In panel (b), we show $R_{xx}(t/t^*)$ confined through a quartic potential, $U(\mathbf{x})=k\,|\mathbf{x}|^4/4$.  
We show the active response for different values of the ratio $v_0^2/T$ comparing ABP and AOUP models. The passive case, obtained for $v_0=0$, is temperature independent since the only dependence on the parameters is contained in $t^*=\sqrt{\gamma}/\sqrt{T\,k}$ and, thus, is simply shown for $T=1$.
Finally, the inset shows the comparison between Eq.~\eqref{eq:Resp_x_overdamped_active} for $T/v_0^2 =10^{-2}$ and Eq.~\eqref{eq:response_active_over_Tzero} for $T=0$, showing the good agreement between the two FDR.
Here, the additional dashed black lines are eye-guides to evidence the different time regimes.
The other parameters are $k=3$, $\gamma=1$, $\tau=1$ and $v_0=1$. 
}
\end{figure}

To check the results also in the overdamped case, we numerically study the response function considering the same confining potentials studied in the underdamped case: i) quadratic potential, $U(\mathbf{x})=k\, |\mathbf{x}|^2/2$ and ii) quartic potential $U(\mathbf{x})=k\, |\mathbf{x}|^4/4$ where $k$ is the potential constant.
In both cases, the cross elements of the response function ($R_{x_i,x_j}$ with $i \neq j$) are zero for the same symmetric arguments already explained for the underdamped dynamics.
The time, $t$, is evaluated in unit of the typical time, $t^*$, that rules the response decay of the passive overdamped system, given by $t^*=\gamma/k$ for the harmonic potential and $t^*=\sqrt{\gamma}/\sqrt{T\,k}$ for the quartic potential.
With this time rescaling, $R_{x_i,x_j}(t)$ does not depend on the model parameters, in the passive case.

In both cases, the response function evaluated numerically from the perturbed dynamics (see definition~\eqref{eq:numericalresponse_u}) is compared with the FDR, Eq.~\eqref{eq:Resp_x_overdamped_active}, showing a good agreement for different values of $v_0^2/T$ both for the AOUP and ABP models. This confirms the validity of our exact relations also in non-equilibrium dynamics.
Fig.~\ref{fig:over}~(a) illustrates the response function in the harmonic case, where the decay is exponential, $R_{x_j x_j}(t) = e^{-t/t^*}$, as analytically predicted in Ref.~\cite{szamel2014self} for the athermal AOUP.
In the harmonic case, we observe that there are no differences between AOUP, ABP, and passive systems. As a consequence, the shape of the active force is irrelevant despite the non-Gaussian form of the active force in the ABP model.
In Fig.~\ref{fig:over}~(b), the response function in the quartic potential case shows a richer behavior.
The rescaled $R_{x_j x_j}(t/t^*)$ has an exponential profile that does not depend on the choice of $k/\gamma$ and $T$.
This profile coincides with the active one, in the equilibrium limit $v_0^2 \ll T$ (shown for $v_0^2/T=10^{-1}$), where AOUP and ABP cannot be distinguished simply because the active force is negligible.
Increasing the ratio $v_0^2/T$, the active response starts decreasing faster even if there are no clear differences between AOUP and ABP models, that appear only for further values of $v_0^2/T$.
In general, the decay of $R_{x_j x_j}(t)$ is faster for the ABP model than the AOUP one, and the difference between the two models increases when $v_0^2/T$ grows. In this regime, the decay is characterized by two distinct time-regimes, as explicitly shown in the inset of Fig.~\ref{fig:over}~(b). As also discussed in~\cite{caprini2020fluctuation} for $T=0$, these two regimes can be easily explained because an active particle (in the large persistence regime, considered here, for $v_0^2 \gg T$) confined in a quartic potential accumulates on a circular crown far from the potential minimum showing pronounced non-gaussianity in the distribution~\cite{caprini2019activity}.
We observe that the formulation~\eqref{eq:Resp_x_overdamped_active} of the FDR reported in this work does not coincide with the recent one, obtained for an AOUP particle with zero solvent-temperature.
Indeed, the FDR for $T=0$, reported in~\cite{caprini2020fluctuation} involves the second derivative of the potential that is not contained in Eq.~\eqref{eq:Resp_x_overdamped_active}.
Moreover, Eq.~\eqref{eq:Resp_x_overdamped_active} is not well-defined at $T=0$ even if can be numerically evaluated for $T$ arbitrarily small.
In the inset of Fig.~\ref{fig:over}~(b), the expression~\eqref{eq:Resp_x_overdamped_active} and the formulation of Ref.~\cite{caprini2020fluctuation} (that for completeness is reported in ~\ref{eq:FDR_Tzero})
reveals a good agreement between the two formulations of the AOUP response function when $T \ll v_0^2$ revealing the convergence of the two generalized FDR in this limit.







\subsubsection{Generalized Virial equation and Equipartition theorem}

In the case of an active particle in the underdamped regime following the dynamics~\eqref{eq:dynamics_activecolloids}, the relation~\eqref{eq:passive_colloidal_equip} (for a passive colloid) turns to be:
\begin{equation}
\label{eq:virial1_active}
 m\langle v_k^2 \rangle + \frac{m}{\gamma}\langle v_j \nabla_{x_k} U \rangle \delta_{jk}  =  T +  \frac{m}{\gamma}\langle v_j \mathrm{f}^a_k  \rangle \delta_{jk} \,,
\end{equation}
where we have reported the relation for $j=k$ for simplicity.
In practice, the interpretation of the terms involved in this equation does not change with respect to Eq.~\eqref{eq:passive_colloidal_equip},
except for the presence of a new term, i.e. $ \langle v_j \mathrm{f}^a_k\rangle$, appearing in the generalized version of the equipartition theorem. This can be easily interpreted as the work done by the active force that is responsible for the increase of the particle kinetic energy.
A similar scenario occurs by generalizing Eq.~\eqref{eq:passive_colloidal_virial} to the active dynamics. In particular, taking $j=k$, we obtain: 
\begin{equation}
\label{eq:virial2_active}
\langle x_j \nabla_{x_k} U \rangle\delta_{jk} = \langle x_j \mathrm{f}^a_k  \rangle \delta_{jk} + m\langle v_k^2 \rangle \,.
\end{equation}
Now, the generalized Virial equation contains a new term that depends on the active force via its correlation with the particle position, which is proportional to minus the swim pressure (See \cite{takatori2014swim,winkler2015virial,marini2017pressure}). The term on the left-hand side of Eq.~\eqref{eq:virial2_active} is proportional to the Virial pressure (as in the case of passive colloids). We remark that, in the active case, the Virial pressure is not simply determined by the kinetic energy but is affected by the swim pressure.

In a similar way, we can apply Eq.~\eqref{eq:generalized_equipartition} to the overdamped active dynamics, Eq.~\eqref{eq:motion}, obtaining a set of relations, that we report for $j=k$, for simplicity:
\begin{equation} 
\label{eq:equation_of_state_active_overdamped}
  \langle x_j \nabla_{x_k} U \rangle\delta_{jk} = \delta_{jk} T  + \frac{1}{\gamma}\langle x_j \mathrm{f}^a_k\rangle \delta_{jk} \,.
\end{equation}
Eq.~\eqref{eq:equation_of_state_active_overdamped} is the equation of state (mesoscopic virial equation) for the active dynamics, that can also be obtained by Eqs.~\eqref{eq:virial1_active} and~\eqref{eq:virial2_active} in the limit $m/\gamma \ll 1$.
Here, the virial pressure is modified by the swim pressure as in Eq.~\eqref{eq:virial2_active} and the kinetic energy has been replaced by the solvent temperature.

\section{Comparison with other versions of the FDR}\label{sec:comparison}

Under very general hypothesis, the response function due to a small perturbation for the general dynamics~\eqref{eq:dynamics} can be expressed in terms of suitable temporal correlations that involves the log-derivative of the steady-state probability distribution, $P_s(\mathbf{x})$,~\cite{marconi2008fluctuation,sarracino2019fluctuation}.
This result has been independently derived by Agarwal~\cite{A72} in the context of stochastic processes and Vulpiani et al.~\cite{FIV90} for chaotic deterministic dynamics, and reads: 
\begin{equation}
\label{eq:Vulpio_formula}
\mathcal{R}_{A, x_j} (t) = -\left\langle  A(t) \frac{d}{d x_j}\log P_s(\mathbf{x}) \biggr|_{s=0}  \right\rangle \,.
\end{equation}
This relation allows us to express the response function in terms of a temporal correlation that has a very simple form.
The application of Eq.~\eqref{eq:Vulpio_formula} does not require the dynamical knowledge of the deterministic or stochastic contributions appearing in Eq.~\eqref{eq:dynamics} since the knowledge of the steady-state distribution is enough to express the generalized FDR.
However, $P_s(\mathbf{x})$ is known for a few cases: i) linear dynamics with additive noise ii) equilibrium dynamics characterized by zero currents. Indeed, in general, when the detailed balance does not hold the distribution is unknown and the use of Eq.~\eqref{eq:Vulpio_formula} requires the numerical calculation of $\frac{d}{dx_j}\log P_s(\mathbf{x})$. Therefore, this relation remains, somehow, an implicit relation.
We remark that, through this approach, one can calculate the response function directly from the experimental data in the absence of perturbation but, also in this case, still requires to recognize the leading variables appearing in the dynamics.
On the contrary, our exact relations~\eqref{eq:FDR_A_general} and~\eqref{eq:FDR_x_general} for a general observable and for $A=x_k$, respectively, reveal also that the response function cannot be easily expressed in the same form of Eq.~\eqref{eq:Vulpio_formula}, i.e. 
\begin{equation*}
\mathcal{R}_{A, x_j} (t) = \left\langle A(t) \, C(0)   \right\rangle \,,
\end{equation*}
except when the detailed balance holds. Indeed, Eqs.~\eqref{eq:FDR_A_general} and~\eqref{eq:FDR_x_general} contain additional terms that cannot easily be recast onto this form. Finding the functional form of $C$ is a problem with the same difficulty of finding the functional form of the steady-state probability distribution of a non-equilibrium system.

Several later formulations of the FDR based on a path-integral approach focused on the importance of the time-reversal symmetry in the different contributions of the response function. For instance, in~\cite{maes}, the response function has been decomposed in terms of an {\it entropic} and a {\it frenetic} contributions. However, this decomposition goes beyond the aim of this study and, in general, cannot be achieved unless one knows the parity under the time-reversal transformation of each variable appearing in the dynamics.
This parity is often unknown, as occurs for the active force appearing in the dynamics~\cite{dabelow2019irreversibility} and could depend on the physical system under consideration.
Our formulation of the FDR does not need this information and is expressed in a simple and compact form.

\section{Conclusions}

In this paper, we have derived a new version of the generalized Fluctuation-Dissipation relations (FDR) that holds both for equilibrium and non-equilibrium dynamics. The advantage of our relations is that they are expressed in a very compact and simple form in terms of time correlations between the observed variable and the force ruling the dynamics of the perturbed variable.
For this reason, our FDR only requires the knowledge of the deterministic forces and the diffusion matrix appearing in the dynamics and does not need the numerical calculation of the steady-state probability distribution, at variance with other approaches.
From our FDR, we have also derived generalized equations that constrain the steady-state (equal-time) correlation functions of the dynamical variables. These equations are interpreted as generalized versions of the Mesoscopic Virial equation and equipartition theorem.

Our general results have been checked in the case of an equilibrium underdamped and overdamped colloidal particle, where our FDR agree with the well-known equilibrium picture.
Finally, we have also applied our relations to a non-equilibrium system of active particles finding the generalized FDR, the Mesoscopic Virial equation and the equipartition relation both for overdamped and underdamped dynamics.
The present study can be easily generalized to the case of many interacting particles, both for the passive and the active case and could be useful to make further advances in the calculation of the transport coefficients in non-equilibrium dynamics, especially in systems of active matter, going beyond the results obtained for active crystals~\cite{caprini2020time} or low densities~\cite{dal2019linear}.

\section*{Acknowledgement}
The author warmly thanks A. Sarracino, A. Puglisi and U. Marini Bettolo Marconi for interesting discussions and
acknowledges support from the MIUR PRIN 2017 Project No. 201798CZLJ

\appendix

\section{Derivation of Eq.~\eqref{eq:FDR_A_general}}\label{app:1}

To derive Eq.~\eqref{eq:FDR_A_general}, we employ a path-integral approach to estimate the probability of the trajectory associated to the unperturbed dynamics~\eqref{eq:dynamics}.
In the following, we use the compact notation, $\underline{\mathbf{x}} = \{\underline{\mathbf{x}}\}^{\mathcal{T}}_{t_0}$, to denote the time-history of the single trajectory between the initial time, $t_0$ , and the final time, $\mathcal{T}$. The
explict introduction of a source of noise in the dynamics, produces a probability, $\mathcal{P}[\mathbf{x}| \mathbf{x}_0]$, of observing a path $\underline{\mathbf{x}}$ given the initial state $\mathbf{x}_0$ . In the following, we consider Gaussian noises, $\boldsymbol{\eta}$, which are entirely specified by mean values and correlations and that satisfies $\eta_i =\sigma_{ij} \xi_j$.
Under these assumptions, the probability of observing the
noise path, $\underline{\boldsymbol{\eta}}$, reads:
\begin{equation}
\label{eq:noise_probability}
\mathcal{P} [\underline{\boldsymbol{\eta}} | \boldsymbol{\eta}_{t_0}] \propto \exp{\left[-\frac{1}{4} \int^{\mathcal{T}} ds\, D^{-1}_{ij} \eta_j(s) \eta_i(s)  \right]} \,,
\end{equation}
where we dropped an irrelevant normalization factor and used the Einstein convention for repeated indices.

Observing that the functional derivative with respect
to the perturbation, $h_j$, is equivalent to the functional derivative with respect to the noise, $\eta_j$, we obtain an expression for the response function starting from its definition~\eqref{eq:response_def}:
\begin{eqnarray}
\mathcal{R}_{A, x_j} (t-s) &&=   \frac{ \delta\langle A(\mathbf{x}^h(t))\rangle}{\delta h_j(s)}\biggr|_{h_n=0}  = \int^t \mathcal{D}[\underline{\boldsymbol{\eta}}] \mathcal{P}[\underline{\boldsymbol{\eta}}|\boldsymbol{\eta}_{t_0}] \frac{\delta}{\delta \eta_j(s)} A(\mathbf{x})=  \\
&&=- \int^t \mathcal{D}[\underline{\boldsymbol{\eta}}]  A(\mathbf{x}) \frac{\delta}{\delta \eta_j(s)} \mathcal{P}[\underline{\boldsymbol{\eta}}|\boldsymbol{\eta}_{t_0}]    \,,
\end{eqnarray}
where in the last equality we have just performed an integration by parts.
Using the expression~\eqref{eq:noise_probability} and performing the derivative, we get:
\begin{equation}
\label{eq:noise_A_resp}
\mathcal{R}_{A, x_j} (t-s) = \frac{1}{2}\langle A(t) D^{-1}_{kj}\eta_k(s)\rangle \,.
\end{equation}
Using the dynamics~\eqref{eq:dynamics}, one can express the noise $\boldsymbol{\eta}$ in terms of the state variables $\mathbf{x}$ through a change of variables, so that, formally, we have $\boldsymbol{\eta} = \boldsymbol{\eta}[\mathbf{x}, \dot{\mathbf{x}}]$.
By replacing this relation into~\eqref{eq:noise_A_resp}, we obtain:
\begin{equation}
\label{eq:app_1}
\mathcal{R}_{A, x_j} (t-s) = \frac{1}{2}\left\langle A(t) D^{-1}_{kj}\left[ \dot{x}_k(s) - \mathcal{F}_k(s) \right]\right\rangle \,.
\end{equation}
Because of the stationarity of the correlations, the following relation holds:
\begin{equation*}
\left\langle A(t) D^{-1}_{kj} \dot{x}_k(s) \right\rangle = \frac{d}{ds}\left\langle A(t) D^{-1}_{kj} x_k(s) \right\rangle = -\frac{d}{dt}\left\langle A(t) D^{-1}_{kj} x_k(s) \right\rangle \,,
\end{equation*}
so that we can express Eq.~\eqref{eq:app_1} as
\begin{equation}
\label{eq:app_general_result}
\mathcal{R}_{A, x_j} (t-s) =- \frac{1}{2}D^{-1}_{kj} ò\left[ \left\langle A(t)  \mathcal{F}_k(s)\right\rangle + \frac{d}{dt}\left\langle A(t) x_k(s) \right\rangle \right] \,,
\end{equation}
that corresponds to Eq.~\eqref{eq:FDR_A_general} after choosing $s=0$.

\subsection{Assuming the detailed balance}

Assuming the equilibrium condition or the detailed balance means the possibility of flipping the time in the temporal correlation appearing in Eq.~\eqref{eq:app_general_result}.
In particular, using this property, we get:
\begin{equation*}
\left\langle A(t) x_k(s) \right\rangle = \pm\left\langle x_k(t) A(s) \right\rangle \,,
\end{equation*}
where the plus or minus sign is needed if the product between $x_k A$ is even or odd under time-reversal transformation, respectively.
In this way, Eq.~\eqref{eq:app_general_result} reads
\begin{eqnarray}
\mathcal{R}_{A, x_j} (t-s) &&= - \frac{1}{2}D^{-1}_{kj} \left[ \left\langle A(t)  \mathcal{F}_k(s)\right\rangle \mp \frac{d}{dt}\left\langle x_k(t) A(s) \right\rangle \right] \nonumber\\
&&=  - \frac{1}{2}D^{-1}_{kj} \left[ \left\langle A(t)  \mathcal{F}_k(s)\right\rangle \mp \left\langle \mathcal{F}_k(t) A(s)  \right\rangle \right] =  - D^{-1}_{kj} \left\langle A(t)  \mathcal{F}_k(s)\right\rangle    \,, 
\end{eqnarray}
where, in the last equality, we have used again the reversibility condition and that $\mathcal{F}_k$ needs to have the same parity of $x_k$ in equilibrium dynamics.
This equilibrium result is in agreement with the other version of the FDR~\cite{marconi2008fluctuation,sarracino2019fluctuation}, given by Eq.~\eqref{eq:Vulpio_formula}.
Indeed, if the detailed balance holds the distribution associated with the dynamics~\eqref{eq:dynamics} is simply:
\begin{equation*}
P_{s}\propto \exp{\left( \int dx_i D^{-1}_{ k i}\mathcal{F}_{k}   \right)} \,.
\end{equation*}

\section{Derivation of Eq.~\eqref{eq:FDR_x_general}}\label{app:B}

To derive Eq.~\eqref{eq:FDR_x_general}, we start fom the formula~\eqref{eq:FDR_A_general}.
Choosing $A(\mathbf{x}(t))=x_k$, we obtain:
\begin{equation}
\label{eq:app_resp_x_general_first}
\mathcal{R}_{x_k, x_j} (t) =- \frac{1}{2}D^{-1}_{mj} ò\left[ \left\langle x_k(t)  \mathcal{F}_m(0)\right\rangle + \frac{d}{dt}\left\langle x_k(t) x_m(0) \right\rangle \right] \,.
\end{equation}
Replacing $d/dt \,x_k$ with the equation of motion~\eqref{eq:dynamics}, we get:
\begin{equation}
\label{eq:app_resp_x_general}
\mathcal{R}_{x_k, x_j} (t) =- \frac{1}{2}D^{-1}_{mj} ò\left[ \left\langle x_k(t)  \mathcal{F}_m(0)\right\rangle + \left\langle \mathcal{F}_k(t) x_m(0) \right\rangle \right] \,,
\end{equation}
which corresponds to the result~\eqref{eq:FDR_x_general}.
We remark that to obtain Eq.\eqref{eq:app_resp_x_general} from Eq.~\eqref{eq:app_resp_x_general_first} we have used the causality condition, such that $\langle \eta_k(t) x_m(0) \rangle=0$. 
This trick can be also used in the more general case $A=A(\mathbf{x}(t))$:
\begin{eqnarray}
\frac{d}{dt}\left\langle A(\mathbf{x}(t)) x_m(0) \right\rangle &&= \left\langle \dot{x}_j(t)\nabla_{x_j} A(\mathbf{x}(t)) \,x_m(0) \right\rangle \nonumber\\
&&= \left\langle \mathcal{F}_j(t)\nabla_{x_j} A(\mathbf{x}(t)) \,x_m(0) \right\rangle + \left\langle \eta_j(t)\nabla_{x_j} A(\mathbf{x}(t)) \,x_m(0) \right\rangle \,.
\end{eqnarray}
Moreover, in this case, the correlation involving the noise does not vanish because the causality condition cannot be applied.
At variance with the specific case reported in Eq.~\eqref{eq:FDR_x_general}, this last general relation is not simply expressed in terms of state variables and, thus, has the same level of complexity as Eq.~\eqref{eq:FDR_A_general}.

\subsection{Assuming the detailed balance}
Even if we have shown the result for a general observable $A$, it is instructive to use the time-reversibility to further manipulate Eq.~\eqref{eq:FDR_x_general}. 
In particular, if the product $\mathcal{F}_j(t) x_k(0)$ is even under time-reversal transformation, the second term in Eq.\eqref{eq:FDR_x_general} becomes
\begin{equation*}
\langle \mathcal{F}_j(t) x_k(0) \rangle = \langle x_k(t) \mathcal{F}_j(0) \rangle \,,
\end{equation*}
while, if the product $\mathcal{F}_j(t) x_k(0)$ is odd, the following relation holds:
\begin{equation*}
\langle \mathcal{F}_j(t) x_k(0) \rangle = -\langle x_k(t) \mathcal{F}_j(0) \rangle \,.
\end{equation*}
Thus, the response matrix can be expressed as:
\begin{equation}
\label{eq:responsematrix_detailedbalance}
\mathcal{R}_{x_j, x_n} (t) =-D_{nk}^{-1}  \left[\langle x_j(t) \mathcal{F}_k(0) \rangle \pm \langle x_k(t) \mathcal{F}_j(0) \rangle \right] \,.
\end{equation}
Further manipulation of this expression can be obtained accounting for the symmetry of the system. For instance, if the equilibrium is guaranteed by a force due to an external potential that depends only on the distance $\sqrt{\sum_j x_j^2}$, the system is invariant for the inversion of each component and $\mathcal{F}_j$ needs to an odd function of $x_j$.
Thus, the non-vanishing elements of the response matrix are those with $j=n$ and $\mathcal{F}_j$ and $\langle x_j(t) \mathcal{F}_k(0) \rangle = \langle x_k(t) \mathcal{F}_j(0) \rangle $.
In this way, Eq.~\eqref{eq:responsematrix_detailedbalance} leads to the well-known equilibrium result.

\section{FDR for zero solvent temperature}\label{eq:FDR_Tzero}
In this Appendix, we report the FDR obtained in the case of overdamped active particles evolving with the AOUP model with vanishing solvent temperature, i.e. the dynamics~\eqref{eq:dynamics_activecolloids} with $T=0$ and active force evolving via Eq.~\eqref{eq:AOUP}.
In this case, the formulation~\eqref{eq:Resp_x_overdamped_active} of the FDR does not hold
since the equation of motion is not of the form~\eqref{eq:dynamics_perturbed}. Indeed, the perturbation, $h$, affects the dynamics of a state variable $x_j$ with a deterministic equation of motion, since the noise appears only in the evolution of the active force.

For completeness, we report the FDR expression derived in~\cite{caprini2020fluctuation}, holding for the athermal AOUP, that has been employed in the inset of Fig.~\ref{fig:over}~(b), for the quartic potential case:
\begin{eqnarray}
\label{eq:response_active_over_Tzero}
2D_a\gamma\mathcal{R}_{x_j x_i} =& \langle x_j(t) \nabla_{x_i}U(0) \rangle + \langle \nabla_{x_i}U(t) x_j(0)  \rangle \\
&+ \tau^2 \langle v_j(t) \nabla_{x_j}\nabla_{x_k} U(0) v_k(0) \rangle + \tau^2 \langle v_k(t) \nabla_{x_i}\nabla_{x_k} U(t) v_j(0) \rangle \,, \nonumber
\end{eqnarray}
where the particle velocity, defined as $v_j = \dot{x}_j$, satisfies the following relation:
\begin{equation*}
\gamma v_j = \mathrm{f}^a_j - \nabla_{x_j} U
\end{equation*}
We also remark that Eq.\eqref{eq:response_active_over_Tzero} depends on the details of the active force. 
In particular, it does not hold for the athermal ABP model, for which explicit generalized FDR have not been derived, for the best of our knowledge.


\section*{References}

\bibliography{Respbib.bib}

\end{document}